\documentclass[natbib,times]{aastex}
\usepackage{spr-astr-addons}
\usepackage{url}
\usepackage{graphicx}

\usepackage{natbib}
\bibpunct{(}{)}{;}{a}{}{,}

\RequirePackage{color}

\newcommand{\emaila}{Wolfgang.Baumjohann@oeaw.ac.at}

\begin{document}

\title{Ultrarelativistic generalized Lorentzian thermodynamics\\ and  the differential cosmic ray energy flux}
\shorttitle{Cosmic ray flux spectra}
\shortauthors{R. A. Treumann and W. Baumjohann}

\author{R. A. Treumann\altaffilmark{1,2}} \and \author{W. Baumjohann\altaffilmark{3}}
\email{\emaila}

\altaffiltext{1}{International Space Science Institute, Bern, Switzerland.}
\altaffiltext{2}{Geophysics, Munich University, Munich, Germany.}
\altaffiltext{3}{Austrian Academy of Sciences, Space Research Institute, Graz, Austria.}

\begin{abstract}
We apply the ultrarelativistic generalized Lorentzian quasi-equilibrium thermodynamic energy distribution to the energy spectrum of galactic cosmic ray fluxes. The inferred power law slopes contain a component which evolves with cosmic ray energy in steps of thirds, resembling the sequence of structure functions in fully developed Kolmogorov turbulence. Within the generalized thermodynamics the chemical potential can be estimated from the deviation of the fluxes at decreasing energy. Both may throw some light on the cosmic ray acceleration mechanism to very high energies.
\end{abstract}


\section{Introduction}

{Half a century ago Stan Olbert\footnote{See the acknowledgement in \citep[at page 2867]{1968vasyliunas}.} introduced an \emph{ad hoc} model distribution function which since became known as $\kappa$-distribution.
It was successfully applied to {formally fitting} moderate-energy electron  \citep[first by][]{1968vasyliunas} and ion fluxes \citep[cf.,][as for examples]{1988christon,1991christon} in space. As a statistical probability distribution it ultimately found various applications \cite[cf.,][for a review]{2013livadiotisSSRv}. In physics, the appearance of $\kappa$ distributions as substitutes for the Maxwell-Boltzmann distribution in the stationary long-term limit \citep{1985hasegawa-PRL,2012yoonPoP,2012yoonSSRv,2012yoonPoP-a} when particles are accelerated in interaction with fluctuating electromagnetic fields or turbulence led to suspect that they obey a deeper physical meaning in quasi-equilibrium statistical mechanics. Such a statistical mechanics, dubbed quasi-equilibrium generalized-Lorentzian statistical mechanics or thermodynamics \citep{1999treumannPS-a,1999treumannPS-b,2008treumannPRL}, was based on the \emph{grand-canonical Gibbs-Lorentzian} probability distribution \citep{2014treumannFP}. The $\kappa$-probability distribution appeared as a version of the generalized Lorentzian phase-space quasi-equilibrium (in the above long-term sense). Its application to relativistic bosons in quantum field theory was given in \citep{2016treumann-epl}. }  

{Generalized energy/momentum space Lorentzians are non-stochastic thermodynamic quasi-equilibrium momentum space distributions in the presence of correlations \citep{1999treumannPS-a,2014treumannFP} provided, for instance, by interaction with a bath of plasma waves and parameterized by the power index $\kappa$. Their non-stochastic poperties have so far prevented construction of a counting statistics. Their consistent correlationless, i.e. the purely stochastic limit $\kappa\to\infty$, is the Boltzmann-Maxwell phase space distribution.}

{The generalized-Lorentzian statistical mechanics strictly satisfies the two \emph{thermodynamic equilibrium} requirements: particle number and energy conservation. Any higher order moments have no physical meaning and thus are of no physical interest. Application to non-equilibrium conditions like conservation laws in fluid mechanics including, for instance, heat flux, require adjusting the power in the energy/momentum distribution. Non-physical higher order moments can be renormalized by adding an exponential truncation factor which cuts the distribution off above some maximum energy. }

{In the present note we apply the \emph{generalized Lorent\-zian equilibrium} thermodynamics to observations of \emph{ultrarelativistic} cosmic ray spectra, interpreting them as thermodynamic quasi-equilibria.}

\section{Ultrarelativistic generalized Lorentzians}
Generalized Lorentzian thermodynamics, for application to cosmic ray fluxes, must be given in ultrarelativistic form, with $\epsilon_\mathbf{p}\approx pc\gg mc^2$ the particle energy. The ultrarelativistic generalized Gibbs-Lorentzian partition function \citep{2014treumannFP} yields the grand-canonical Gibbs-Lorentzian phase-space energy distribution (for our purposes suppressing the exponential cut-off which in cosmic rays comes into play only above ``ankle'' energies)
\begin{equation}
w_{i\kappa}(\epsilon_i)=A\big\{1+(\epsilon_i-\mu)/\kappa T\big\}^{-\kappa-r}
\end{equation}
of finding a particle at thermodynamic temperature $T$ (in energy units) in state $i$. The transition is made from $w_{i\kappa}(\epsilon_i)\to f_\kappa(\mathbf{p})$ to the momentum distribution function $f_\kappa(\mathbf{p})$, normalizing it to the number density $N/V$, with volume $V$, according to $N/V=(2\pi\hbar)^3\int d^3p\,f_\kappa(\mathbf{p})$. In the ultrarelativistic case this yields the generalized Lorentzian momentum space distribution
\begin{equation}
f_\kappa^\mathit{ur}(\mathbf{p})=\frac{N\lambda^3}{V\kappa^3}\frac{\big\{1+(pc-\mu)/(\kappa T)\big\}^{-\kappa-r}}{\mathrm{B}(3,\kappa+r-3)}
\end{equation}
{which is a \emph{grand canonical thermodynamic equilibrium} distribution, the momentum space particle density. As required in equilibrium it conserves particle number and energy.} $\mu$ is the chemical potential related to particle number, $\lambda=2\pi\hbar c/T$ is the ultrarelativistic thermal de Broglie length, and $\mathrm{B}(a,b)$ is the Beta function. {Here $T$ is not a so-called `equivalent' temperature, it is the \emph{exact physical thermodynamic temperature} \citep{2014treumannFP}, the inverse of which is the partial derivative of the thermodynamic energy with respect to the entropy.} 

The constant $r$ in the exponent is fixed from the {thermodynamic requirement} \citep{2014treumannFP} that the mean ultrarelativistic energy  $U_\mathrm{ur}$ of an ideal gas is related to particle number {$N$ and thermodynamic temperature $T$} via the ultrarelativistic ideal gas equation $U_\mathrm{ur}=3 NT${, or written in terms of the ultrarelativistic pressure $P_\mathrm{ur}V=3NT$}. Averaging the particle energy $\epsilon_\mathbf{p}$ yields straightforwardly
\begin{equation}
U_\mathrm{ur}=\frac{3\kappa}{\kappa+r-4}N T
\end{equation}
which identifies $r=4$, with ultrarelativistic Lorentzian 
\begin{equation}\label{eq4}
f_\kappa^\mathit{ur}(\mathbf{p})=\frac{N\lambda^3}{2V\kappa^4}\frac{(\kappa+3)!}{\kappa!\big\{1+(pc-\mu)/(\kappa T)\big\}^{\kappa+4}}
\end{equation}
Transition to field theory, if necessary, is achieved by putting $p=\hbar k$. The ultrarelativistic generalized-Lorentzian energy space density becomes
\begin{equation}
f_\kappa^\mathrm{ur}(\epsilon)=\frac{\epsilon^2}{c^3}f_\kappa^\mathrm{ur}({p})\Big|_{pc=\epsilon}
\end{equation}

\begin{figure}[t!]
\centerline{\includegraphics[width=0.5\textwidth,clip=]{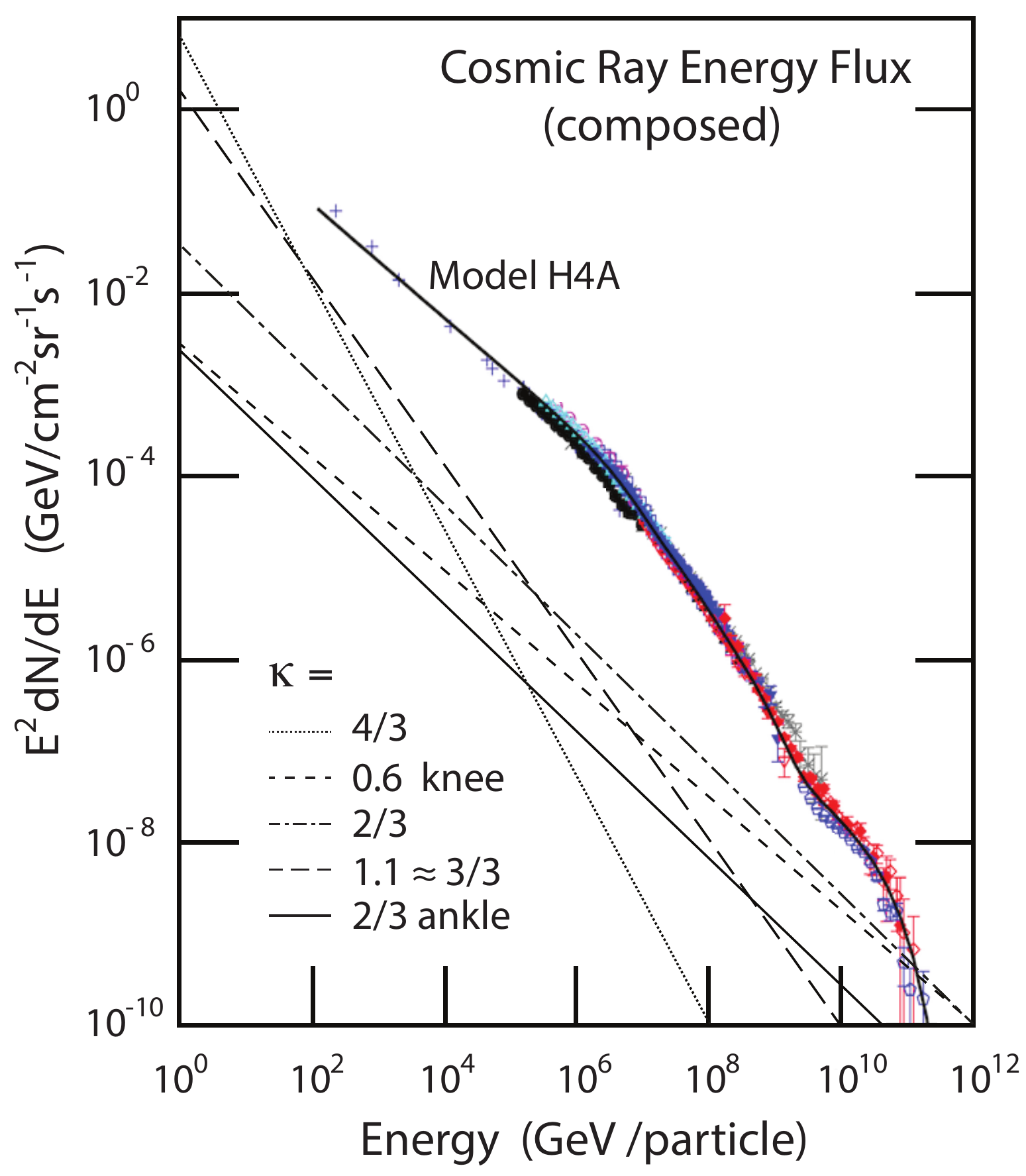}}
\caption{The (publicly available) cosmic ray energy spectrum used in this paper. The data are composites from many instruments. The solid curve is a model fit H4A.  The various differently dashed straight lines indicate the approximate slopes of power laws read from the data if interpreting them in terms of the generalized Lorentzian thermodynamics.} \label{epl-CR-fig1}
\end{figure}

These distributions have some interesting properties \citep{2008treumannPRL}. Firstly, the chemical potential must be negative: $\mu<0$. For $pc<-\mu=|\mu|$ the momentum space density flattens out \citep{2004treumannPoP} towards low $p$ to become $f_\kappa^\mathrm{ur}(p)=$ const, while the energy distribution $f_\kappa^\mathrm{ur}(\epsilon)\propto \epsilon^2$ assumes a parabolic dependence. It maximizes at energy 
\begin{equation}
\epsilon_m= \frac{\kappa T}{1+\kappa/2}\bigg(1+\frac{|\mu|}{\kappa T}\bigg)
\end{equation}
This is an important point, because inspection of measured momentum fluxes at small momenta $mc\ll p<-\mu/c$, if available and flattening out, allows for the determination of $|\mu|$, while measuring $\epsilon_m$ provides the ratio of chemical potential to temperature. On the other hand, for $pc\gg-\mu$ the phase space distributions become simple power laws in either case
\begin{equation}
f_\kappa^\mathrm{ur}(p)\propto \Big(\frac{pc}{\kappa T}\Big)^{-(\kappa+4)},\quad f_\kappa^\mathrm{ur}(\epsilon)\propto\Big(\frac{\epsilon}{\kappa T}\Big)^{-(\kappa+2)} 
\end{equation}
Measuring the power law slope of the distribution at high momentum or energy immediately determines the power $\kappa$ of the generalized Lorentzian. In the following we will make use of these ways in examining the observed cosmic ray energy spectrum. 

\section{The cosmic ray energy spectrum}
Such observations usually measure number fluxes in dependence on energy: 
\begin{equation}\label{eq7}
\frac{cN}{V}=c\int \mathrm{d}\epsilon\,\mathrm{d}\Omega\, f_\kappa^\mathrm{ur}(\epsilon) \Longrightarrow \frac{c}{V}\frac{\mathrm{d}N}{\mathrm{d}\epsilon\,\mathrm{d}\Omega}=cf_\kappa^\mathrm{ur}(\epsilon)
\end{equation}
directly yielding the slope of the energy distribution from the observations. Once the slope has been determined experimentally, the ratio of chemical potential to temperature is inferred from the spectral maximum of the fluxes $f^\mathrm{ur}_\kappa(\epsilon_m)$ in Eq. (\ref{eq7}).  With known $\kappa$ and $|\mu|$ known from the momentum distribution, the temperature $T$ can then be obtained. It is simple matter to show that with $f_0\equiv \big[V/N(2\pi\hbar c)^3\big]\,f^\mathrm{ur}_\kappa\big(p\ll|\mu|/c\big)$, the temperature follows as
\begin{equation}
\kappa T= \bigg[\frac{(\kappa+3)!}{(\epsilon_mf_0)\kappa!}\bigg]^{(\kappa+2)/(3\kappa+5)}\bigg(1+\frac{\kappa}{2}\bigg)^{-1/(3\kappa+5)}
\end{equation}
A similar expression holds for the energy distribution at maximum energy $f_\kappa^\mathrm{ur}(\epsilon_m)$. If $|\mu|$ cannot be determined from either of them, $T^{-1}=(\partial U/\partial S)\big|_V$ as derivative of energy with respect to entropy $S$ must be calculated from the theoretical expression for the entropy \citep{2014treumannFP}. Working out this programme completes the quasi-equilibrium generalized-Lorentzian thermodynamics of cosmic ray fluxes. In the cosmic ray energy fluxes, $\epsilon_m$ and $f_0$ are not directly observable thus inhibiting its completion without adding further information. Instead, inferring the power law index $\kappa$ from published cosmic ray fluxes Fig. \ref{epl-CR-fig1} is of sufficient interest as it contains the hidden microscopic physics of the acceleration.

The above distribution holds if the observer is placed \emph{inside} the generation region. For cosmic ray observations this is not the case, however, as cosmic rays readily escape from their source and propagate out into space where they become further accelerated. One thus expects that just the power law tail, possibly with some weak deviations from power law at low energies, becomes visible, indicating flattening of the distribution. 

The theoretical distribution $f^\mathrm{ur}_\kappa(\epsilon)$ at energies above $\epsilon>-\mu$ is monotonous. Deviations from monotony indicate effects not covered by one {single} function. Also for stationary spectra one can neglect the effect of time dispersion such that the observed long-term spectra can be taken as averaging out all contributions of time variability of the cosmic ray sources.

\section{Power law index}
The observed spectrum of galactic cosmic ray particle fluxes (taken from publicly available spectra reproduced in Fig. \ref{epl-CR-fig1}) is not uniform in energy. The fluxes exhibit different energy ranges of different well-distinguishable slopes $s$. Galactic cosmic rays have composite spectra, as they do not consist of one single particle species. Rather they are composed of different particle components coming from a variety of sources and being subject to primary and secondary acceleration processes \citep{2002schlickeiser-book}. 

These slopes can be read from Fig. \ref{epl-CR-fig2} \citep{2005antoni,2008amenomori,2011apelPRL,2011apel-a,2013aartsenPR,2016aartsenPRL,2008abbasiPRL,2008abrahamPRL,2008aharonian,2015ivanov-a,2015ivanov-b,2015valino,2015tinyakov,2016aab} and are given in Table \ref{tab1} for the different energy ranges in three forms, as slope $s$, $\kappa =s-2$, and in the fractional form $\kappa$ closest to the inferred index $s-2$. (One may note that $\kappa=s-2$ is the genuine power law index of the generalized Lorentzian, with the number $2=r-2$ the thermodynamically required constant addition to warrant energy conservation and adjusting for the equation of state. Thus it is $\kappa$, not $s$,  which contains the hidden physics.) 

{The indices $s$ are approximate in the sense of the precision of the data scatter on the log-log scale with an estimated uncertainty in slope $s$ of $\lesssim 0.1$. However, the three determined ranges are suffciently far apart from each other that this uncertainty is of little importance when considering the slopes on an equal theoretical basis. The last column in Table \ref{tab1} is the closest fractional approximation of the determined $\kappa$ index within this uncertainty of the measurement. These fractions are sufficiently close to the measurements and sufficiently far apart from each other to justify the fractional approximation. (For instance, $\kappa\sim 0.7$ is close to $0.66\dots$ and $\sim0.4$ away from the next $\kappa\lesssim 1.1$.)}

\begin{table}[t]
\caption{{Slope $s$, index $\kappa=s-2$, and closest fractal $\kappa$}}
\begin{center}
\begin{tabular}{c||c|c|c}
$\epsilon$ [TeV]&$s$&$\kappa=s-2$&$\kappa$\\[0.5ex]
\hline\\[-1.5ex]
$<10^3$ &$\sim2.7$ &$\sim0.7$ &$\sim\frac{2}{3}$\\[0.5ex]
$\lesssim10^4$ &$\sim2.6$&$\sim0.6$& ``knee''\\[0.5ex]
$10^4 - 10^5$&$\sim3.1$&$\sim1.1$&$\sim\frac{3}{3}$\\[0.5ex]
$10^5 - 10^6$&$\sim 3.4$&$\sim\frac{10}{3}$&$\sim\frac{4}{3}$\\[0.5ex]
$>10^6$&$\sim2.6$&$\sim0.6$&``ankle''
\end{tabular}
\end{center}
\label{tab1}
\end{table}%

\section{Discussion} 
\begin{figure}[t!]
\centerline{\includegraphics[width=0.5\textwidth,clip=]{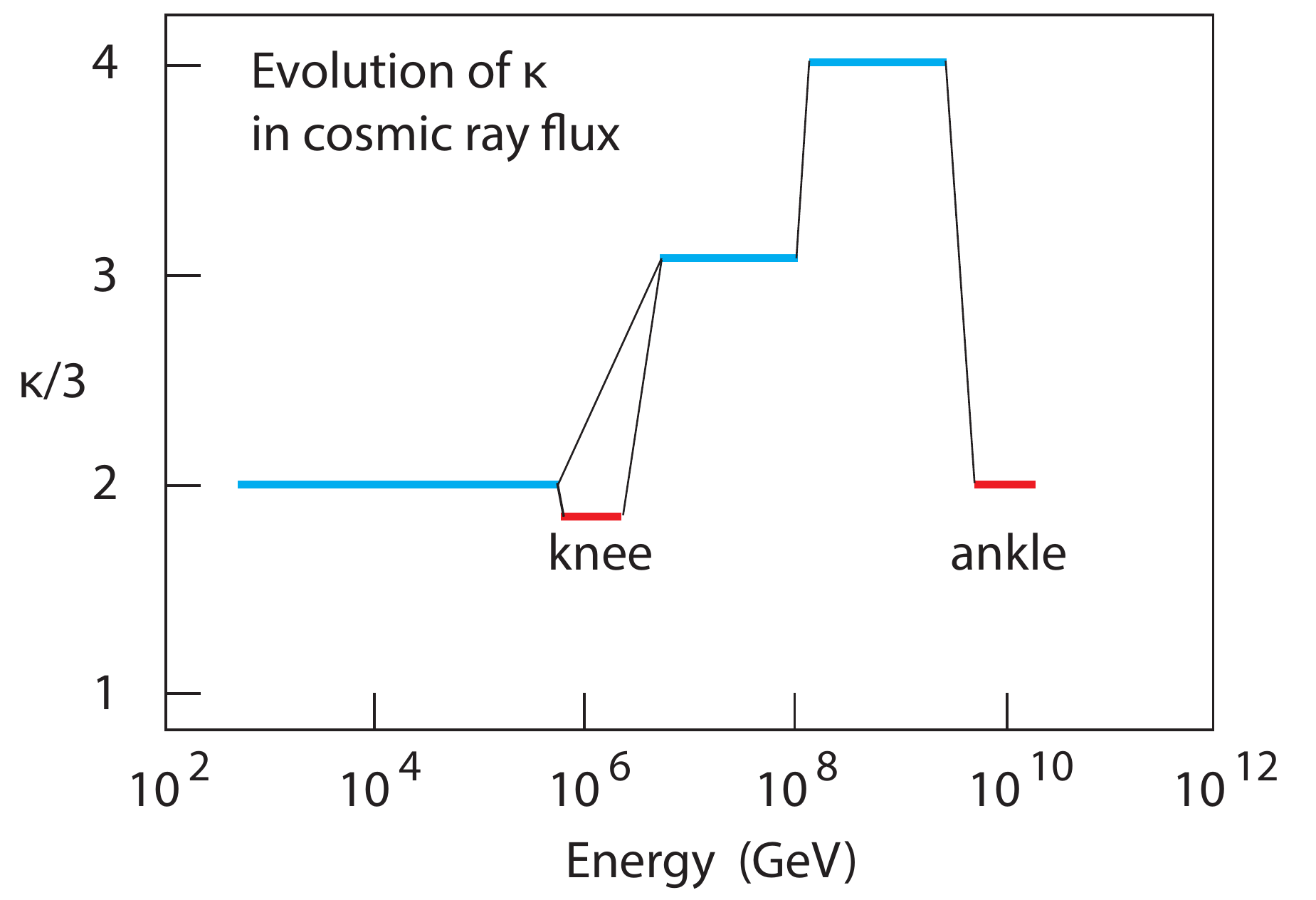} }
\caption{Evolution of the $\kappa$ index of an ultrarelativistic generalized Lorentzian distribution applied to the available cosmic ray fluxes as function of energy \citep{2016beatty,2016ChPhC}. With increasing energy the $\kappa$ index evolves in three consecutive steps if approximately one third, indicating three different stages of acceleration of the cosmic rays which may or may not be related. If they are related they would indicate that with higher energy stepwise higher turbulent structure functions become involved into the diffusive Fermi acceleration process. The regions of the ``knee'' and ``ankle'' have low $\kappa$ values, suggesting  large chemical potentials.} \label{epl-CR-fig2}
\end{figure}

The list of slopes suggests that the index $\kappa$ of the ultrarelativistic quasi-equilibrium Lorentzian thermodynamics applied to the galactic cosmic ray spectrum (excluding the bump of the ``knee'') changes gradually in steps of thirds with energy before, at the ``ankle'', returning to its initial slope $\kappa\sim\frac{2}{3}$.  On excluding ``knee'' and ``ankle'', one thus has the sequence 
\begin{equation}
\kappa_n=n/3, \quad n=2,3,4
\end{equation}
which is graphically shown in Fig. \ref{epl-CR-fig2}. {The changes in slope with energy indicate that the cosmic ray energy fluxes are not determined by one single process only over more than 10 orders of magnitude in energy. It is known that they are composed of particles of different chemical composition while coming from different sources. However, the evolution of $\kappa$ in well defined quantized steps of thirds indicates some inherent process hidden in the evolution of the spectra. It consists\footnote{Closer inspection, in particular of the vicinity of the ``knee'' but also in other sections of the spectrum allows to define a more elaborate sequence of changing slopes, a power law ``fine structure'',  which we just note here but do not investigate anyhow closer.} of -- at least -- three different stages. This fact extracted from the observations should throw light on the acceleration process of charged particles up to the highest cosmic ray energies of $>10^7$ TeV, while it is most interesting that this internal spectral structure is uncovered by \emph{interpreting} the cosmic ray spectra as generalized Lorentzian thermodynamic quasi-equilibrium spectra.}

This scaling in $\kappa$ reminds of Kolmogorov's \citep{1962kolmogorov,1991kolmogorov-a,1991kolmogorov-b} scaling of the turbulent structure functions \citep{1975monin-book,1978frisch} in fully developed inertial range turbulence whose exponents $\zeta_n$ are given in similar sequential order 
\begin{equation}
\zeta_n=n/3, \quad n=2,3,\dots
\end{equation}
{In a picture where the power law spectrum of cosmic rays is progressively generated by the interaction between the charged particle component of the magnetic turbulence as the main feature of diffusive Fermi acceleration, this similarity between the quantised steps of the spectral slope and the structure functions suggests that at increasing energies stepwise higher order turbulent structure functions start contributing to or even dominating the diffusive acceleration of the charged galactic cosmic ray particles to ever higher energies. As particles get relativistically heavier with increasing energy, Fermi acceleration becomes more efficient. One might speculate that this leads to the involvement of higher order turbulent structures.}

\section{Remarks and conclusions}
Independent of the above, in the generalized Lorentzian thermodynamic quasi-equilibrium interpretation both the ``knee'' and ``ankle'' indicate the possible working of additional processes. Flattening of the spectrum at the ``knee'' suggests a shift in energy space towards higher energies, which occurs over roughly half an order of magnitude before the steeper slope belonging to the next third in $\kappa$ takes over. In this interpretation the flattening indicates the start of a new Lorentzian quasi-equilibrium that takes over at higher energies. At ``knee'' energy the effect of the chemical potential of the superimposed Lorentzian would be non-negligible. The energy of the ``knee'' is then identified with the transition to the new higher energy Lorentzian equilibrium and thus a change in chemical potential. This implies the onset of a different process at ``knee'' energies, for instance the above mentioned inclusion of the next-higher order turbulent structure function.

The meaning of such a high energy chemical potential is difficult to understand. In the region far below ``knee''-energy the chemical potential can be read from the deviation of the power law at energies around $\epsilon\lesssim10$ GeV {(data not included here in Fig. \ref{epl-CR-fig1})} and might thus be related to the $\epsilon=1$ GeV rest mass energy of protons. In an analogous interpretation, a chemical potential at ``knee'' energy would suggest the presence of some unknown hypothetical particle of `very high energy/rest mass' the order of say $m\gtrsim10$ TeV$/c^2$, which should then have been generated in the primary cosmic ray production process in the galactic sources. Afterwards they would have passed the turbulent acceleration process to become further accelerated. Because of their large mass, diffusive acceleration would be rather efficient. 

Similarly, from the sudden flattening of the cosmic ray fluxes at ``ankle''-energies, a chemical potential of the order of $|\mu|\lesssim(10^4-10^5)$ TeV would be inferred. Tempted to assume indication of another high energy particle, the inferred rest mass should be of the order of, say, $m>(10^2-10^3)$ TeV$/c^2$. Nucleons of such rest mass energies are so far unknown in our accessible Universe. 

{The accepted interpretation of the cosmic ray flux spectra \citep[cf., e.g.,][]{2002schlickeiser-book} and their ``knee'' and ``ankle'' sections is in terms of superposition of acceleration spectra of the various known chemical particle components generated in AGNs, stars, supernovae and shocks \citep{2013balogh-book,2011treumannAARv}. These particles undergo many acceleration cycles in diffusive Fermi acceleration of the various heavy charged known elemental nucleons in turbulent magnetic fields \citep{2002schlickeiser-book} being small-angle scattered and slowly pushed up in energy until reaching the quasi-equilibrium state in the generalized Lorentzian thermodynamics.}

This slow though efficient diffusive accelecration process is by now quite well established. It involves distributed magnetic fields and many of such magnetic scattering centers located within a large spatial volume. It has by now been sufficiently well developed being in its ultimate completion phase while including a variety of anomalous processes, for instance astrophysical turbulence where the turbulent mechanical energy is ultimately dissipated in electric current filaments \citep{2015treumann-aarv} on leptonic scales by spontaneous collisionless reconnection and as well by ultrarelativistic collisionless shock waves \citep{2011treumannAARv}, which generates the necessary diffusivities in energy and momentum space.  

{The conventional interpretation is satisfactory because it involves only well-established processes, even though some details are not completely understood yet. As an observational fact resulting from the application of the generalized Lorentzian thermodynamics to the cosmic ray energy spectrum, the sequential fractional order of the $\kappa$ parameter extracted from the slope is worth further investigation of its physical meaning. }

\small{\acknowledgments
This work was part of a Visiting Scientist Programme at the International Space Science Institute Bern executed by RT in 2007. We acknowledge the interest of the ISSI Directorate and hospitality of the ISSI staff. We thank the ISSI system administrator S. Saliba for technical support and the librarians Andrea Fischer and Irmela Schweizer for access to the library and literature.}

\nocite{*}
\bibliographystyle{spr-mp-nameyear-cnd}
\bibliography{cr-aa}

\end{document}